\documentclass[aps,pre,twocolumn,floats,superscriptaddress]{revtex4}

\usepackage{latexsym,amsmath}
\usepackage{amsbsy}
\usepackage[pdftex]{graphicx}
\usepackage{amssymb}
\usepackage{epstopdf}
\usepackage[usenames]{color}
\usepackage{float}
\usepackage{hyperref}
\usepackage{graphicx}
\usepackage[normalem]{ulem}
\usepackage{mathtools}
\usepackage{xcolor}
\definecolor{g}{rgb}{0.,0.,0.7} 
\definecolor{gnew}{rgb}{0.,0.4,0.4} 
\definecolor{R}{rgb}{0.5,0.0,0.9} 

\begin{document}

\title{Predictability of missing links in complex networks}

\author{Guillermo Garc\'{\i}a-P\'{e}rez}\thanks{\noindent The two authors contributed equally to this work.}
\affiliation{QTF Centre of Excellence, Turku Centre for Quantum Physics, Department of Physics and Astronomy, University
of Turku, FI-20014 Turun Yliopisto, Finland}
\affiliation{Complex Systems Research Group, Department of Mathematics and Statistics, University
of Turku, FI-20014 Turun Yliopisto, Finland}
\author{Roya Aliakbarisani}\thanks{\noindent The two authors contributed equally to this work.}
\affiliation{Faculty of Computer Engineering, K.~N.~Toosi University of Technology, Tehran 1631714191, Iran}
\author{Abdorasoul Ghasemi}
\affiliation{Faculty of Computer Engineering, K.~N.~Toosi University of Technology, Tehran 1631714191, Iran}
\author{M. \'Angeles Serrano}
\affiliation{Departament de F\'{\i}sica de la Mat\`eria Condensada, Universitat de Barcelona, Mart\'{\i} i Franqu\`es 1, 08028 Barcelona, Spain}
\affiliation{Universitat de Barcelona Institute of Complex Systems (UBICS), Universitat de Barcelona, Barcelona, Spain}
\affiliation{ICREA, Pg.~Llu\'{\i}s Companys 23, E-08010 Barcelona, Spain}

\begin{abstract}

Predicting missing links in real networks is an important problem in network science to which considerable efforts have been devoted, giving as a result a vast plethora of link prediction methods in the literature. In this work, we take a different point of view on the problem and study the theoretical limitations to the predictability of missing links. In particular, we hypothesise that there is an irreducible uncertainty in link prediction on real networks as a consequence of the random nature of their formation process. By considering ensembles defined by well-known network models, we prove analytically that even the best possible link prediction method for an ensemble, given by the ranking of the ensemble connection probabilities, yields a limited precision. This result suggests a theoretical limitation to the predictability of links in real complex networks. Finally, we show that connection probabilities inferred by fitting network models to real networks allow to estimate an upper-bound to the predictability of missing links, and we further propose a method to approximate such bound from incomplete instances of real-world networks. 
\end{abstract}

\maketitle

\section{Introduction}

Limits of predictability, the degree to which a system's state can be correctly forecasted, have been explored in different contexts, including weather and climate~\cite{Slingo:2011}, human mobility~\cite{Song:2010}, and biological evolution~\cite{Nosil:2018}. One of the causes that undermines perfect predictability in these systems---apart from lack of information, observational errors, or variability in their environmental conditions---can be found in the inherent randomness of some of the processes and phenomena that shape their organization and behavior.

In complex networks~\cite{Newman:2010}, randomness not only dominates the dynamical interactions between the states of nodes in many dynamical processes~\cite{Barrat:2008}, which limits the ability to predict specific configurations of dynamical states at any given time~\cite{Radicchi:2018}, but also link formation. The structure of complex networks is far from deterministic and can be modeled in a stochastic framework where the likelihood of links to exist is characterized probabilistically. The set of link probabilities defines a network ensemble, that can be studied to gain insight into some specific network that can be considered to be an instance of such ensemble instead of an independent entity. 

This uncertainty in the likelihood of connections represents an intrinsic feature of networks that affects the predictability of their structure~\cite{Lu:2015}. Link prediction methods~\cite{Lu:2011,Liben-Nowell:2007} are able to give information about missing or future interactions in networks by exploiting the non-trivial regularities in their organization---heterogeneous degree distributions, high levels of clustering, degree-degree correlations, communities---, at the local or at the global level. Different link prediction methods typically give different results on the same network and, although some methods may comparatively perform better than others, it is not clear which is the best precision that can be achieved.

Our hypothesis is that, regardless of how much link prediction methods improve, they will always present an irreducible lack of accuracy in real networks as a consequence of the random nature of link-formation processes. In this work, we address the question of what is the best possible link prediction method for a given network ensemble, and what is its maximal expected precision. We find that, if a network is an instance of a network ensemble, the best link prediction method  simply corresponds to ranking the likelihoods of missing links according to the corresponding connection probabilities given by the ensemble model, which suggests a theoretical limitation to the predictability of links in real complex networks. We name this strategy the Optimal Strategy (OS) for link prediction. We first prove that the OS gives the best possible link prediction method in networks belonging to model ensembles. Then, we show that inferred connection probabilities in well-fitted network models allow to estimate the limit to the predictability of missing links in real networks. 

\section{The Optimal Strategy for a random graph ensemble}
Our goal in this section is to calculate the upper-bound for the maximal accuracy attainable by any link prediction method in networks whose connectivity is described in probabilistic terms. We compute the accuracy in terms of precision, which measures the fraction of correctly predicted links as compared to the total number of missing links. We consider ensembles $\mathcal{E}_N$ defined as sets of networks $G$ of $N$ nodes generated by assigning undirected links between pairs of nodes $i$ and $j$ with independent pairwise probabilities $\lbrace p_{ij} \rbrace$, where the indices run from 1 to $N$. Therefore, every network in the ensemble is weighted by a probability $P(G)$, given by
\begin{equation}
P(G) = \prod \limits_{i<j} p_{ij}^{a_{ij}} (1 - p_{ij})^{1-a_{ij}},
\end{equation}
where the adjacency-matrix entries $\lbrace a_{ij}\rbrace$ take the value $a_{ij} = 1$ if $i$ and $j$ are connected or $a_{ij} = 0$ otherwise. Therefore, $\sum_G P(G) = 1$.

Given a graph $G$, we construct {\it observed} graphs $G_{\mathrm{obs}}$ by removing a fraction $q$ of links selected randomly. A link prediction method $\mathcal{M}$ can be regarded as a map $G_{\mathrm{obs}} \mapsto G_{\mathrm{inf}}$, or $G_{\mathrm{inf}} = \mathcal{M}(G_{\mathrm{obs}})$, that produces a new graph $G_{\mathrm{inf}}$ by adding predicted links to $G_{\mathrm{obs}}$ such that both $G$ and $G_{\mathrm{inf}}$ have the same number of links (we assume that the number of missing links is known to the link prediction method). Let $Q \equiv Q(G,G_{\mathrm{obs}},G_{\mathrm{inf}})$ be the precision of the prediction, defined as the fraction of predicted links that belong to $G$. Thus, if $G_{\mathrm{inf}} = G$, $Q = 1$. 

For a given ensemble $\mathcal{E}_N$, the optimal link prediction method, that is, the one maximizing the expected precision $\langle Q \rangle$ in link prediction experiments over ensemble instances, is the one that generates $G_{\mathrm{inf}}$ from $G_{\mathrm{obs}}$ by adding the links according to the connection probabilities $\lbrace p_{ij} \rbrace$ ranked in decreasing order. To prove it, we compute the expected precision as 
\begin{equation}
\begin{aligned}
\langle Q \rangle &= \sum \limits_{G} \sum \limits_{G_{\mathrm{obs}}} P(G,G_{\mathrm{obs}}) Q(G,G_{\mathrm{obs}},G_{\mathrm{inf}}) \\
&= \sum \limits_{G_{\mathrm{obs}}} P(G_{\mathrm{obs}}) \bar{Q}(G_{\mathrm{obs}},G_{\mathrm{inf}})\label{eq:avg_q},
\end{aligned}
\end{equation}
where $P(G,G_{\mathrm{obs}})$ is the joint probability distribution for a graph $G$ in the ensemble and an observed graph $G_{\mathrm{obs}}$. We have defined $\bar{Q}(G_{\mathrm{obs}},G_{\mathrm{inf}})$ as the expected precision of the link prediction method over all possible original graphs yielding $G_{\mathrm{obs}}$ upon random removal of links. Notice that, in the summation over original graphs $G$, we must take into account that $G_{\mathrm{inf}}$ is independent of $G$; this is the crucial property leading to a limit to the predictability of missing links. Indeed, since more than one original network $G$ can generate the same $G_{\mathrm{obs}}$ upon random link removal, it is impossible for any link prediction method, which maps $G_{\mathrm{obs}}$ into \textit{the same} inferred network $G_{\mathrm{inf}}$ regardless of the original $G$, to give a perfect prediction.

We now ask which is the link prediction method $\mathcal{M}$ that maximizes the expected precision for a given observed graph. After some calculations (see Appendix \ref{app:a}), we find that the average over graphs of the ensemble of the precision given $G_{\mathrm{obs}}$ can be expressed as the scalar product of two vectors,
\begin{equation}
\bar{Q}(G_{\mathrm{obs}},G_{\mathrm{inf}})=\frac{1}{L} \mathbf{\bar{v}} \cdot \textbf{v}_{\mathrm{inf}},
\end{equation}
where $L$ is the number of missing links in $G_{\mathrm{obs}}$ with respect to $G$. The dimension of the vectors equals the number $M$ of potential links (disconnected pairs of nodes) in $G_{\mathrm{obs}}$. If we enumerate the ensemble probabilities as $\lbrace p_{l}\rbrace$, we can write $\mathbf{\bar{v}} = \left(\frac{qp_{1}}{1-p_{1} + qp_{1}}, \ldots, \frac{qp_{M}}{1-p_{M} + qp_{M}}\right)$, where each entry gives the probability that the corresponding link, missing in $G_{\mathrm{obs}}$, is in $G$. The entries in vector $\textbf{v}_{\mathrm{inf}} = \left(a_{1}^{\mathrm{inf}}, \ldots, a_{M}^{\mathrm{inf}}\right)$ correspond to the adjacency-matrix elements of $G_{\mathrm{inf}}$ for the set of potential links of $G_{\mathrm{obs}}$.

The precision is then maximized when the vectors are maximally aligned. By definition of the link prediction method $ \mathcal{M}$, $\textbf{v}_{\mathrm{inf}}$ is a vector containing $L$ values equal to one, while the rest of entries are zero. Clearly, the maximum value for the precision will be obtained if its non-zero entries are placed at the same positions where the $L$ largest components of $\mathbf{\bar{v}}$ are located. Therefore, given that $\mathbf{\bar{v}}_i > \mathbf{\bar{v}}_j \Leftrightarrow p_{i} > p_{j}$, the best link prediction method, the Optimal Strategy (OS), is the one that adds the $L$ missing links according to the highest connection probabilities in the ensemble. Moreover, we see that the expected optimal precision for the observed graph is given by the mean of the $L$ largest components of $\mathbf{\bar{v}}$, that is, by
\begin{equation}
\bar{Q}(G_{\mathrm{obs}},G_{\mathrm{inf}}^\mathrm{opt}) = \frac{1}{L} \sum \limits_{l=1}^{L} \frac{qp_{l}}{1-p_{l} + qp_{l}},
\label{maxav}
\end{equation}
where now index $l$ runs over the set of potential links of $G_{\mathrm{obs}}$ with the corresponding probabilities ordered in decreasing order.

Finally, from the expression above, we observe that the expected optimal precision goes to zero as the number of missing links decreases and it converges to the mean of the top-$L_0$ connection probabilities in the ensemble~---where $L_0$ stands for the number of links in the original graph $G$---, when $q$ is maximal. In fact, the \textit{precision curve is an increasing function of the number of removed links}. This apparently counter-intuitive result stems from the fact that, as the fraction of missing links $q$ increases, the ratio of missing links over potential links in $G_{\mathrm{obs}}$, given by $\frac{qL_0}{N(N-1)/2-(1-q)L_0}$, increases and so the probability of missing an actually missing link decreases. Therefore, the statistical power of the method, {\it i.e.} the probability that the prediction of a missing link is correct, increases with $q$.

\begin{figure*}[t]
\includegraphics[width=\textwidth]{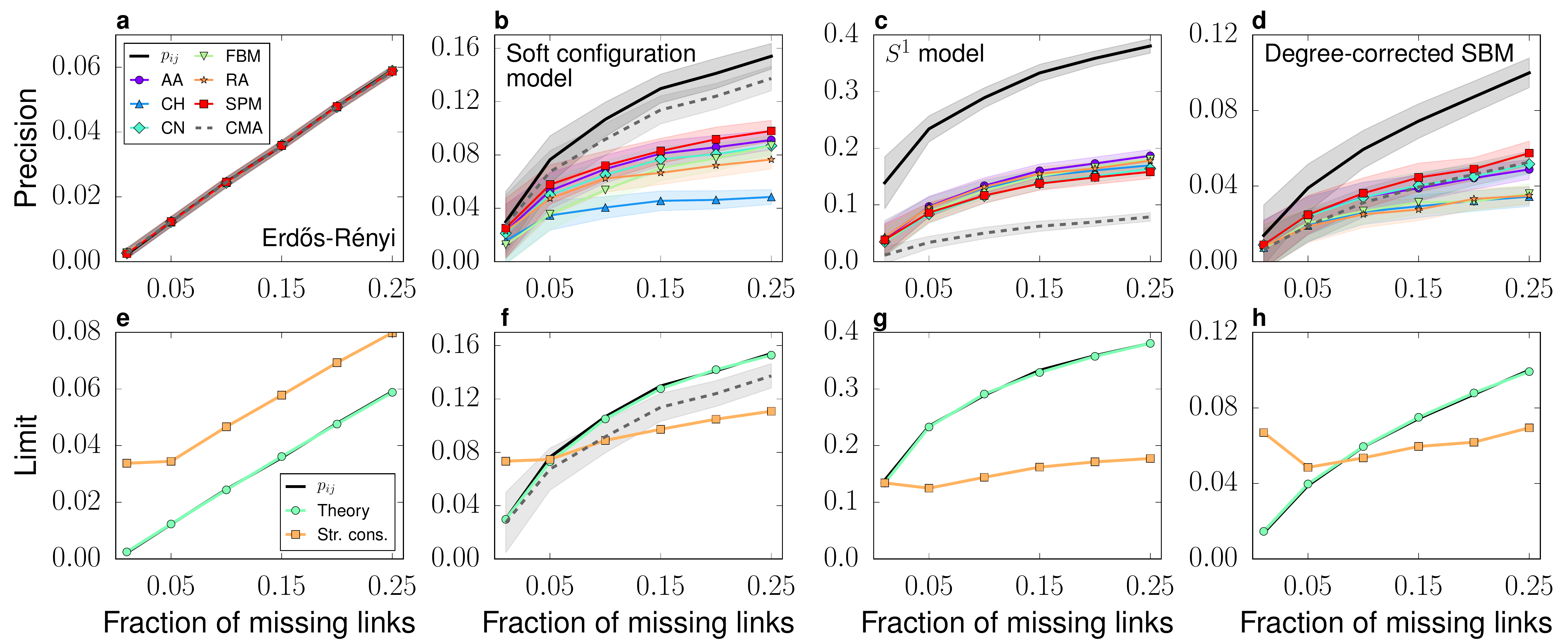}
\caption{\label{fig:fig1}\textbf{a}-\textbf{d}: Precision as a function of the fraction of missing links for different link prediction methods on networks of four different ensembles. For each ensemble and value of the fraction of missing links $q$, we generated 10 networks $G$ and, for each one of them, we generated 100 incomplete networks $G_{\mathrm{obs}}$ on which the link prediction methods were applied. The shaded areas represent the standard deviation of the results. The $p_{ij}$ curves correspond to the theoretical limit computed through numerical simulations which implement the Optimal Strategy. In all cases, we have set $N = 1000$ nodes. Also, except for the ER ensemble, we have set $\langle k \rangle = 10$ and power-law degree distributions with exponent $\gamma = 2.5$. \textbf{a}: ER networks with $p = 0.2$. \textbf{b}: Soft Configuration Model. \textbf{c}: $\mathbb{S}^1$ model with $\beta = 1.5$. \textbf{d}: Degree-corrected Stochastic Block Model with $\lambda = 0.5$ and 7 equiprobable blocks. \textbf{e}-\textbf{f}: Comparison between the OS predictability curve (analytical estimation) and the Structural Consistency index ~\cite{Lu:2015} (details in Appendix \ref{app:sci}).}
\end{figure*}

\subsection{The Optimal Strategy on network ensembles defines the OS predictability curve}
We compared the predictability curve given by the Optimal Strategy (OS predictability curve), computed through numerical simulations (see analytical estimation in Appendix \ref{app:tos}), against the precisions of several link prediction methods on different network ensembles. We consider six widely applied link prediction methods. Four of them---Common Neighbors (CN) ~\cite{Newman2001CN}, Adamic-Adar (AA)~\cite{Adamic:2003}, Resource Allocation (RA)~\cite{Zhou:2009RA}, and Cannistraci-Hebb (CH)~\cite{Cannistraci:2017CH}---exploit local connectivity patterns, while the other two---Structural Perturbation Method (SPM)~\cite{Lu:2015} and Fast probability Block Model (FBM)~\cite{liu2013correlations}---are global (see Appendix \ref{app:b} for details). 

First, we considered the variant of the Erd\H{o}s-R\'enyi (ER) model where all pairs of nodes have the same connection probability $p$~\cite{Erdos:1959,Gilbert:1959}. We observe that link prediction in ER networks is insensitive to the link prediction method used. This result is easily understood from our theoretical analysis, as all possible distributions of the $L$ values equal to one among the (otherwise zero) different components of vector $\textbf{v}_{\mathrm{inf}}$ yield the same scalar product $\mathbf{\bar{v}} \cdot \textbf{v}_{\mathrm{inf}}/L$ and, hence, the same precision. Thus, all methods must give as a result the same precision curve as the Optimal Strategy, which is supported by numerical evidence in Fig.~\ref{fig:fig1}\textbf{a}, where we report the precision as a function of the fraction of missing links for the different link prediction methods. Hence, we can now claim from strong theoretical grounds that ER networks are unpredictable, as conjectured in Ref.~\cite{Lu:2015}. The curves also show that the link prediction accuracy in ER networks is very low, in accordance with previous reports~\cite{Lu:2009,Lu:2015}. 

We also considered the soft Configuration Model (sCM)~\cite{Park:2003}, producing maximally random graphs with a given expected degree sequence, the $\mathbb{S}^{1}$ model~\cite{Serrano:2008ga}, producing maximally random geometric graphs with given expected degree sequence and level of clustering, and the degree-corrected Stochastic Block Model (dc-SBM)~\cite{Karrer:2011}, a generalized block model that accounts for heterogeneities in the degrees to generate networks with given mesoscopic structure, see details in Appendix \ref{app:c}. The results are shown in Fig.~\ref{fig:fig1}\textbf{b}, Fig.~\ref{fig:fig1}\textbf{c}, Fig.~\ref{fig:fig1}\textbf{d}, respectively. As expected, the Optimal Strategy ---the best possible method--- gives the best results in all cases. All ensemble models show link prediction accuracies significantly above those for the ER ensemble, being the $\mathbb{S}^{1}$ ensemble the one with the highest predictability and, at the same time, the one in which link prediction methods perform worse.

\begin{figure*}[t]
\includegraphics[width=\textwidth]{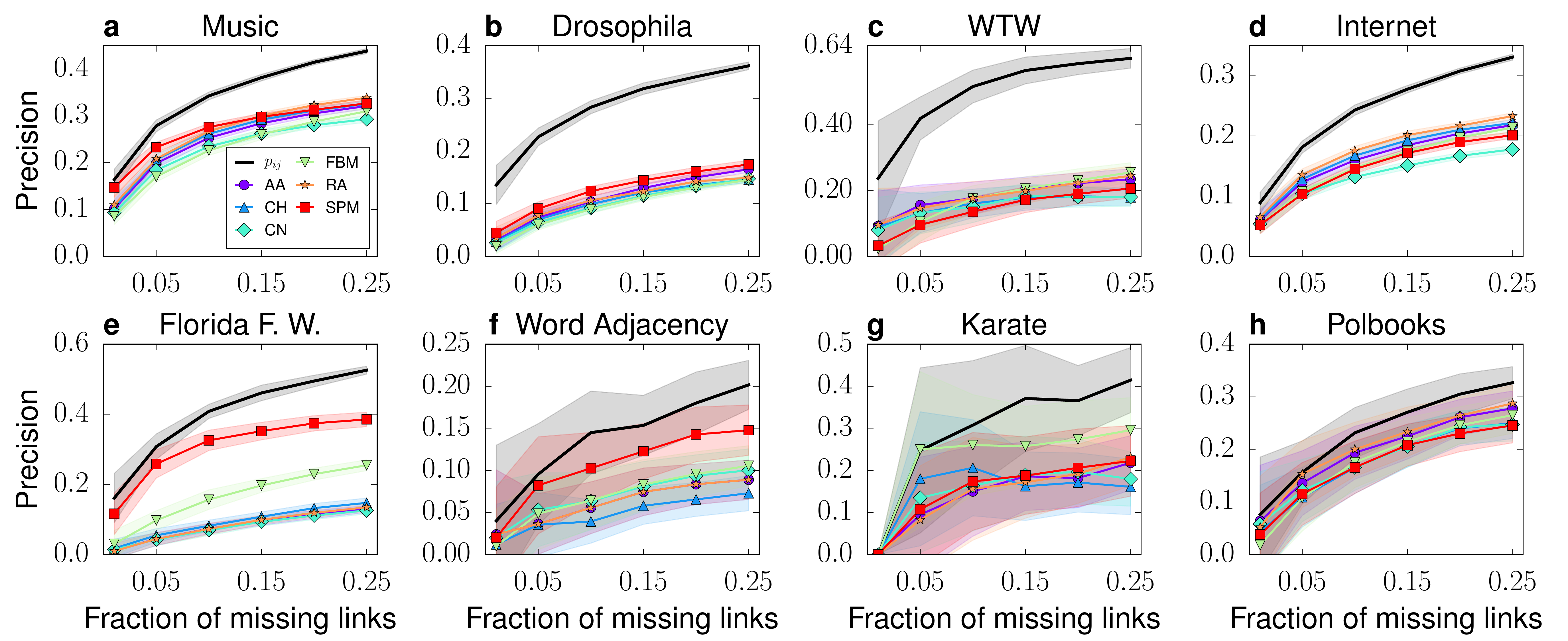}
\caption{\label{fig:fig2}Precision as a function of the fraction of missing links for different link prediction methods on eight real-world networks. For each network and value of the fraction of missing links $q$, we generated 100 incomplete networks $G_{\mathrm{obs}}$ on which we applied the link prediction methods. The $p_{ij}$ curves correspond to the precisions given by the simulations of the Optimal Strategy using the ranking of the inferred probabilities. \textbf{a}-\textbf{d}: Using the $\mathbb{S}^{1}$ model. \textbf{e}-\textbf{h}: Using the Dc-SBM model. In all plots, the shaded areas represent the standard deviation of the results, typically larger for smaller networks.}
\end{figure*}

Our analysis emphasises the importance of identifying correctly the ensemble to which a given network belongs for an accurate prediction of missing links. To further illustrate this point, we infer the ensemble connection probabilities in the soft CM from observed, incomplete networks and then use the resulting ranking as a link prediction method, which we name the Configuration Model Assumption (CMA). These connection probabilities are easy to estimate. In the soft CM, each node $i$ is assigned an expected degree which coincides approximately with the resulting degree $k_{i}$ obtained in realizations of the model. Hence, after randomly removing a fraction $q$ of links from an ensemble network, we expect the observed degree of every node to become $k_{i}^{\mathrm{obs}} \approx (1-q) k_{i}$. Thus, given the observed degrees $k_{i}^{\mathrm{obs}}$ in the incomplete graph, we can estimate the original-network degrees and approximate the connection probabilities accordingly using their definition, Eq.~\eqref{eq:conn_prob_CM} (see Eq.~\eqref{eq:conn_prob_CM_proxy}). We then use the inferred probabilities as scores in link prediction experiments on synthetic networks belonging to all the considered ensembles. As the results in Fig.~\ref{fig:fig1}\textbf{b}-\textbf{d} show, the CMA method works extremely well for networks belonging to the soft CM ensemble, achieving a precision curve higher than any other link prediction method, nearly matching the theoretical maximum given by the OS predictability curve. However, the same link prediction method fails when used on completely different networks, like the $\mathbb{S}^{1}$-model ensemble networks. Interestingly, the results for CMA are comparable to other link prediction methods on dc-SBM networks, mainly due to their explicit resemblance with the assumptions in the sCM model (see Appendix \ref{app:c} for details).

Finally, we compared the OS predictability curve with the structural consistency index, see Fig.~\ref{fig:fig1}\textbf{e}-\textbf{h}. Notice that the structural consistency index, introduced in Ref.~\cite{Lu:2015}ã is not a link prediction method but it was proposed to estimate the link predictability of a network based on the assumption that removing a small subset of links at random from the given network does not change its structural features (further details in Appendix \ref{app:sci}). However, our results show that the Structural Consistency index gives bounds that are impossible to achieve (see, for instance, the results for the ER ensemble and for low values of $q$ in the sCM and the dc-SBM ensembles). Furthermore, it underestimates limits to link predictability that are actually surpassed by some of the methods used, like CMA in the sCM ensemble.

\section{Limits to link prediction in real networks}
Taken together, the results in the previous section support the claim that identifying the model that describes accurately the connectivity structure of a real network would give the best link prediction precision by applying the OS, as long as the underlying ensemble connection probabilities can be inferred from the observed graph. 

\subsection{The OS predictability curve in real networks}
In general, the inference of the ensemble connection probabilities from an observed real graph considered as incomplete is a very difficult problem for models other than the soft CM. Nevertheless, we can still evaluate the predictability curve for real data using information given by the original graph. The main idea is to infer the ensemble probabilities from the original network, before any links have been removed, and to use them to apply the Optimal Strategy on link prediction experiments. The resulting precisions hence indicate the limits of an ideal model-based link prediction strategy, that is, in which the ensemble probabilities can be inferred from the incomplete network.

In this subsection, we apply the aforementioned approach on eight different real networks (see Appendix \ref{app:d} for details). We infer the connection probabilities of the most suitable ensemble model (the one that best reproduces their topologies) before links are randomly removed. Finally, we compare the inferred OS predictability curve, obtained by applying the Optimal Strategy using the inferred ensemble probabilities, with the results given by link prediction methods as a function of the number of missing links. The comparison is shown in Fig.~\ref{fig:fig2}. Notice that the OS predictability curve, in addition to indicating the potential of a model-based link prediction approach, can be used to benchmark currently existing link prediction methods, as their precisions can now be compared against a theoretical upper-bound---namely, the precision of the Optimal Strategy using the probabilities of an ensemble of very similar networks to the real one (given by the inferred probabilities) averaged over random link removals.

The Music, Drosophila, WTW, and Internet networks are well described by the $\mathbb{S}^{1}$ model, due to their heterogeneous degree distributions and high levels of clustering. These networks were embedded in the underlying geometry assumed in the model by finding the parameters that maximise the likelihood for the real graphs to be generated by the model, following the same approach as in Refs.~\cite{boguna:2010, papadopoulos:2015}. Once the angular positions of the nodes in the underlying one-dimensional sphere, or circle, and the hidden degrees are found, the $\mathbb{S}^{1}$-model connection probabilities (see Eq.~(\ref{eq:probability_of_connection_S1}) in Appendix \ref{app:c}) between all pairs of nodes define an ensemble of networks which are similar to the real one. We use these probabilities to compute the OS predictability curves shown in Fig.~\ref{fig:fig2}\textbf{a}-\textbf{d}, which lay well above the precisions obtained by other link prediction methods.

A similar result is observed on the four datasets well described by the dc-SBM, as depicted in Fig.~\ref{fig:fig2}\textbf{e}-\textbf{h}, where we show the results for the Florida Food Web, Word Adjacency, Karate, and Polbooks networks. To compute the connection probabilities of a given network, it is fitted to a dc-SBM to find its community structure using a statistical inference and a Monte Carlo sampling~\cite{riolo:2017}. This procedure computes the number of groups, $K$, and the group assignment, $g$, for the network. Then, the connection probability of every pair of nodes is computed by Eqs.~\eqref{eq:Tetha_dc-SBM} and~\eqref{eq:Conn_Pro_dc-SBM}.
Results in Fig.~\ref{fig:fig2}\textbf{e}-\textbf{h} show that, even if fluctuations are more important than in the previous scenario, the precisions obtained by the different link prediction methods are still lower than the OS predictability curve.

\subsection{Approximating the OS predictability curve from an observed network}\label{sec:approx_OScurve}
The OS predictability curve presented in the last subsection measures how predictable a network is when assumed to belong to some specific model. However, its inference presents an evident difficulty as it requires knowledge of the original network, which is obviously inaccessible ---as it is to be predicted--- to compute the ensemble probabilities $p_{ij}$. To overcome this issue, we propose a method to estimate the OS predictability curve directly from the observed network structure by computing a set of connection probabilities that approximate those in the original ensemble.

\begin{figure}[t]
\includegraphics[width=0.9\columnwidth]{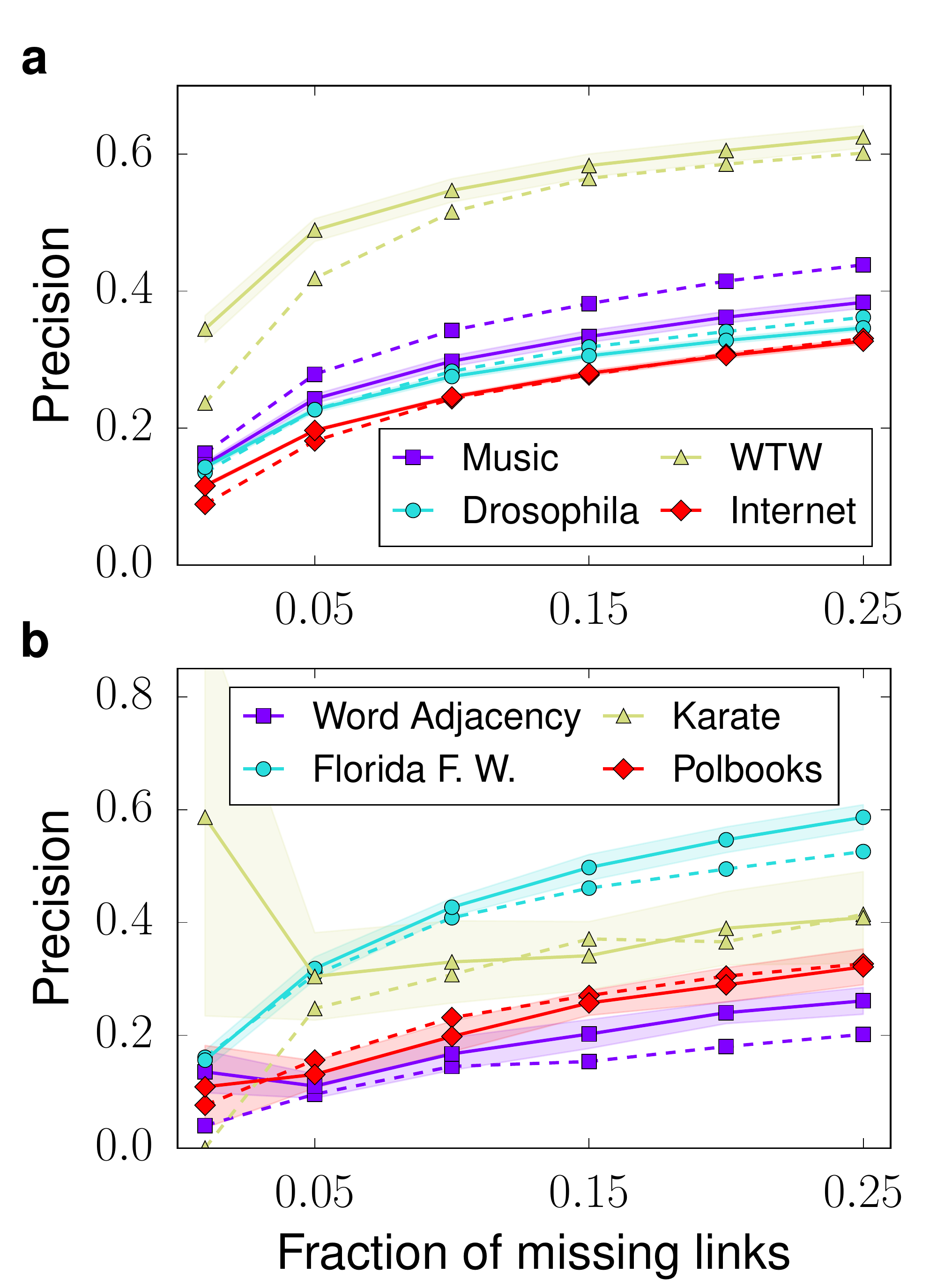}
\caption{\label{fig:fig3}Inference of predictability on eight different real-world networks. The dashed curves show the mean precisions given by the $p_{ij}$, that is, the OS predictability curves, as in Fig.~\ref{fig:fig2}. For each network, we considered 10 incomplete networks with $q_0 = 0.1$, and for each incomplete network, we computed its inferred predictability. The solid curves and shaded areas represent the average and standard deviation of such estimations over the 10 networks.}
\end{figure}

Suppose that network $G_{\mathrm{obs}}$ has been generated by removing a fraction $q_{0}$ of links from an original network $G$. We select the most suitable probabilistic network model (the one that best reproduces the topological features of $G_{\mathrm{obs}}$) and fit it to $G_{\mathrm{obs}}$ to obtain the set of connection probabilities $p_{ij}^{\mathrm{obs}}$. Let $E_{\mathrm{obs}}$ be the number of links in $G_{\mathrm{obs}}$. Since there is a fraction $q_0$ of missing links from $G$, the expected number of links in the complete graph is $E = E_{\mathrm{obs}}/(1-q_0)$. Hence, the number of missing links when a fraction $q$ of links is removed from $G$ and a new graph $\tilde{G}$ is produced is $\tilde{L} = q E_{\mathrm{obs}}/(1-q_0)$.

Next, we rank all links in $G_{\mathrm{obs}}$ in decreasing order according to their connection probabilities $p_{ij}^{\mathrm{obs}}$, that we relabel as $p_{ij}^{\mathrm{obs}} \leftrightarrow p_{l}$, such that $p_{l} > p_{l+1} ,\, \forall l$. Now, the method proceeds by a cumulative sequential computation using the ordered list of probabilities of the expected number of non-links of $\tilde{G}$, $H$, and the expected number of non-links of $\tilde{G}$ that would exist in $G$, $T$. When $H \approx \tilde{L}$ (that is, after predicting the top-ranked $\tilde{L}$ links) the expected precision can be estimated as $\langle Q \rangle = T/H$. Hence, after initializing $H$ and $T$ to zero, we visit every pair of nodes $l = 1, \ldots$ and compute their corresponding increments. Two different situations need to be considered differently:

\begin{itemize}
\item[1.] The two nodes in pair $l$ are connected in $G_{\mathrm{obs}}$. In this case, the link must surely exist in $G$. Therefore, in the ensemble of $\tilde{G}$ networks, the link does not exist (and counts as a correct prediction) with probability $q$, so every time one such link is visited, we must update $T_{\mathrm{new}}=T_{\mathrm{old}}+q $ and $H_{\mathrm{new}}=H_{\mathrm{old}}+q$.

\item[2.] The two nodes in pair $l$ are not connected in $G_{\mathrm{obs}}$. There are two possible reasons for the link not to be observed:
\begin{itemize}
\item[\textit{a.}] The link belongs to $G$, but has been removed from it with probability $q$ when producing $\tilde{G}$. The probability that the link is in the original network is $q_0 p_{l}/(1 - p_{l} + q_0 p_{l})$, so that the probability for it not to belong to $\tilde{G}$ is
\begin{equation}
\mathrm{P} \left( \tilde{a}_l = 0, a_l = 1 | a_{l}^{\mathrm{obs}} = 0 \right) = q \frac{q_0 p_{l}}{1 - p_{l} + q_0 p_{l}}.
\end{equation}

\item[\textit{b.}] The link does not belong to $G$, and therefore it cannot exist in $\tilde{G}$. Since the probability that the link does not exist in $G$ is $(1 - p_{l})/(1 - p_{l} + q_0 p_{l})$ the corresponding probability simply reads
\begin{equation}
\mathrm{P} \left( \tilde{a}_l = 0, a_l = 0 | a_{l}^{\mathrm{obs}} = 0 \right) = \frac{1 - p_{l}}{1 - p_{l} + q_0 p_{l}}.
\end{equation}
\end{itemize}
With these two results, we can readily update $T$ and $H$. Since $T$ accounts for the expected number of correct predictions, only case \textit{a.}~contributes, that is,
\begin{equation}
T_{\mathrm{new}}=T_{\mathrm{old}} + q \frac{q_0 p_{l}}{1 - p_{l} + q_0 p_{l}}.
\end{equation}
As for $H$, both cases contribute, and so
\begin{equation}
H_{\mathrm{new}}=H_{\mathrm{old}} + \frac{1 + (qq_0 - 1) p_{l}}{1 - p_{l} + q_0 p_{l}}.
\end{equation}
\end{itemize}

The results of the inferred OS predictability curve as compared with the original one are shown in Fig.~\ref{fig:fig3} for the real networks and the corresponding ensemble network models reported in Fig.~\ref{fig:fig2}. In all networks, the quality of the inferences is very good, both for the $\mathbb{S}^1$ network model ensemble and for the dc-SBM ensemble.

\section{Discussion} 
Link prediction in real networks remains a major challenge. A clear indicator is given by the current prediction accuracy of the methods in experimental tests where a part of the links is randomly removed, with precisions typically far from its absolute maximum, even for the best methods. Part of this seemingly poor performance is explained by the intrinsic unpredictability of networks, whose links are formed following processes that can be mimicked by stochastic connectivity rules determining the likelihood of interactions. Our probabilistic approach to the predictability problem makes sense as far as this assumption is fulfilled.

We have proven here that the Optimal Strategy for link prediction on networks belonging to some model ensemble simply corresponds to ranking the likelihood of missing links according to the connection probabilities given by the model. Our numerical simulations support our theoretical assertions, as the Optimal Strategy outperforms all the link prediction methods used on different network ensembles. This implies that identifying the model that best describes the connectivity of a given incomplete network and inferring the ensemble connection probabilities generating the complete network would yield the best link prediction accuracies. We have also proven this claim by designing such link prediction strategy for soft Configuration Model networks, the CMA method, which gives, by far, the best predictions on such graphs, nearly reaching the theoretical maximum. Since the Configuration Model misses several key properties of real networks, like the high level of clustering, we do not expect the CMA method to perform well in real situations. However, our results serve as a proof of principle motivating to pursue a similar line of model-based link prediction methods with some more realistic network models, like the $\mathbb{S}^1$ and the dc-SBM. Furthermore, besides the insights that our approach provides to the link prediction problem, the precision of the Optimal Strategy yields a novel indicator of the inherent predictability of network models.

In real networks, we propose a method to assess their predictability based on the assumption that they are well described by probabilistic network models. Hence, by inferring the corresponding model ensemble probabilities and by measuring the precision of the Optimal Strategy with them, we obtain the OS predictability curve. This curve can be used as a benchmark to assess the goodness of link prediction methods, as it allows for their performances to be contrasted against the best possible performance over classes of networks which are statistically similar to the one under study. The inference of the OS predictability curve is, however, a difficult task in real networks. On the one hand, its reliability is subject to the congruency between the network and the probabilistic model that best describes the network structure. Typically, a network model can describe correctly the observed connectivity structure of a real system only to a limited extent. On the other hand, it may happen that some particular link prediction method, more tailored for a single network, yields a better result than the Optimal Strategy. Yet, in terms of the ensemble, such a method would be overfitted and perform worse on average on the set of similar networks defined by the same set of $p_{ij}$. This has clear implications, for instance, in the prediction of missing links as future events in time-evolving networks. A link prediction method that is overfitted to specific realizations, like present network snapshots, will certainly fail more easily in foreseeing future connections.

A different issue is that, in situations in which one may want to assess to what extent a given real network, regarded as incomplete, can be predicted, the ensemble probabilities cannot be directly inferred from the original network, as it is unknown. Given that it is in general very difficult to infer the original ensemble probabilities from the incomplete network---which could be further used for actual link prediction---for models other than the sCM, we propose a method to approximate the OS predictability curve, based on probabilities calculated from the observed network, with good accuracy. We remark that a good approximation of the OS predictability curve is not a guarantee that the calculated probabilities are accurate enough to apply the Optimal Strategy as an efficient link prediction method. This is, for instance, the case of the $\mathbb{S}^1$ ensemble in the real-network experiments shown in Fig.~\ref{fig:fig2}, for which the Optimal Strategy works well when using the inferred probabilities of the complete network but gives bad results (not shown) when using the ones  calculated from the incomplete networks. The reason for this phenomenon can be understood from the description of the method presented in Section \ref{sec:approx_OScurve}. In the approximation of the OS predictability curve, only the highest numerical values of the connection probabilities are used, without any mention whatsoever to the pair of nodes they refer to. Hence, as long as the distribution of the values of the highest probabilities is not drastically perturbed by the link removal---that is, if the highest values of the probabilities inferred on the original and incomplete networks exhibit similar distributions---the OS predictability curve can be estimated, even if the specific probabilities corresponding to the removed edges change considerably and, as a result, they do not enable a good link prediction.

\section{Acknowledgements} 
We thank M.~E.~J. Newman for kindly sharing with us the code to estimate the probabilities of connection in the degree corrected stochastic block model. We thank Mari\'an Bogu\~{n}\'a for helpful discussions. G.~G.-P. acknowledges financial support from the Academy of Finland via the Centre of Excellence program (Project no.~312058 as well as Project no.~287750), and from the emmy.network foundation under the aegis of the Fondation de Luxembourg. R.~A. acknowledges the support of the international affairs of K. N. Toosi University of Technology and M.~A.~S. and the University of Barcelona for their support during her visit. M.~A.~S acknowledges support from a James S. McDonnell Foundation Scholar Award in Complex Systems; Ministerio de Ciencia, Innovaci\'on y Universidades of Spain project no. FIS2016-76830-C2-2-P (AEI/FEDER, UE); and the project {\it Mapping Big Data Systems: embedding large complex networks in low-dimensional hidden metric spaces} -- Ayudas Fundaci\'on BBVA a Equipos de Investigaci\'on Cient\'{\i}fica 2017.

\appendix

\section{Details of the derivation of the optimal prediction for a graph ensemble} \label{app:a}

We simplify the notation by enumerating all potential links (disconnected pairs of nodes) in $G_{\mathrm{obs}}$, such that their ensemble probabilities can be writen as $p_{l}$, where the index $l$ runs from 1 to the number of pontential links $M$. Given the corresponding adjacency matrix elements $\lbrace a_{l}\rbrace_{1 \leq l \leq M}$ of $G$, for every potential link $l$ $P(a_{l}=1|a_{l}^{\mathrm{obs}}=0) = P(a_{l}=1,a_{l}^{\mathrm{obs}}=0)/P(a_{l}^{\mathrm{obs}}=0) = qp_{l}/(1-p_{l} + qp_{l})$, where we have used Bayes' rule. Then, the probability for any graph $G$ compatible with the observed graph $G_{\mathrm{obs}}$, $P(G|G_{\mathrm{obs}})$, can be expressed in terms of the set of pairs as
\begin{equation}
P(G|G_{\mathrm{obs}}) = \prod \limits_{l=1}^{M} \frac{(1-p_{l})^{1-a_{l}}(qp_{l})^{a_{l}}}{1-p_{l} + qp_{l}}.
\end{equation}
Let us furthermore define the vector $\textbf{v} = (a_{1}, \ldots, a_{M})$, which characterizes the set of potential links in $G$, and the analogous vector $\textbf{v}_{\mathrm{inf}} = (a_{1}^{\mathrm{inf}}, \ldots, a_{M}^{\mathrm{inf}})$ for $G_{\mathrm{inf}}$. With these two vectors, we can now express the precision as
\begin{equation}
Q(G,G_{\mathrm{obs}},G_{\mathrm{inf}}) = \frac{1}{L} \textbf{v} \cdot \textbf{v}_{\mathrm{inf}},
\end{equation}
where $L = \sum_{l} a_{l}$ is the number of missing links in $G_{\mathrm{obs}}$ with respect to $G$. Hence, we can express $\bar{Q}(G_{\mathrm{obs}}, G_{\mathrm{inf}})$ in Eq.~\eqref{eq:avg_q} as
\begin{equation}
\begin{aligned}[c]
&\bar{Q}(G_{\mathrm{obs}},G_{\mathrm{inf}})  = \\
&= \sum \limits_{G | G_{\mathrm{obs}} \in \mathcal{S}(G)} P(G|G_{\mathrm{obs}}) Q(G,G_{\mathrm{obs}},G_{\mathrm{inf}}) \\
&= \sum \limits_{G | G_{\mathrm{obs}} \in \mathcal{S}(G)} \prod \limits_{l=1}^{M} \frac{(1-p_{l})^{1-a_{l}}(qp_{l})^{a_{l}}}{1-p_{l} + qp_{l}} \frac{1}{L} \textbf{v} \cdot \textbf{v}_{\mathrm{inf}}\\
 &=\left( \sum \limits_{G | G_{\mathrm{obs}} \in \mathcal{S}(G)} \prod \limits_{l=1}^{M} \frac{(1-p_{l})^{1-a_{l}}(qp_{l})^{a_{l}}}{1-p_{l} + qp_{l}}  \textbf{v} \right) \cdot \frac{\textbf{v}_{\mathrm{inf}}}{L}\label{eq:dot_prod},
\end{aligned}
\end{equation}
where $\mathcal{S}(G)$ stands for the set of subgraphs of $G$. In the above calculation, we have used the linearity of the scalar product and neglected the fluctuations in the number of missing links (we assume that all original graphs generating the observed graph upon random link removal with probability $q$ have approximately the same number of links, $L = q \sum_i p_i$, with the sum now taken over all pairs of nodes). One could actually give an exact result, with no assumptions or approximations about the number of missing links, by defining the precision as the fraction of inferred links actually belonging to the original graph (true positive rate). In that case, both $L$ and the inferred vector are the outcome of the link prediction method and the above expression is exact. 

Let us call the vector within the parenthesis in the equation above $\mathbf{\bar{v}}$. Its $n$-th component can be computed as
\begin{equation}
\begin{aligned}
&\sum \limits_{G | G_{\mathrm{obs}} \in \mathcal{S}(G)} \prod \limits_{l=1}^{M} \frac{(1-p_{l})^{1-a_{l}}(qp_{l})^{a_{l}}}{1-p_{l} + qp_{l}}  a_{n}= \\
&= \sum \limits_{a_{1}=0}^{1} \cdots \sum \limits_{a_{M}=0}^{1} \prod \limits_{l=1}^{M} \frac{(1-p_{l})^{1-a_{l}}(qp_{l})^{a_{l}}}{1-p_{l} + qp_{l}}  a_{n}= \\
&= \frac{qp_{n}}{1-p_{n} + qp_{n}} \prod \limits_{l \neq n} \left( \frac{qp_{l}}{1-p_{l} + qp_{l}} + \frac{1- p_{l}}{1-p_{l} + qp_{l}} \right) \\
&= \frac{qp_{n}}{1-p_{n} + qp_{n}}.
\end{aligned}
\end{equation}
Hence, $\mathbf{\bar{v}} = \left(\frac{qp_{1}}{1-p_{1} + qp_{1}}, \ldots, \frac{qp_{M}}{1-p_{M} + qp_{M}}\right)$. 

\section{link prediction methods used in this work} \label{app:b}
\begin{itemize}
\item[] \textbf{Common Neighbours} (CN)~\cite{Newman2001CN}: Local link prediction method in which the pairs of unconnected nodes with more neighbours in common are more likely to be connected by a link. Hence, links are ranked according to the score given by
\begin{equation}
CN(x,y)= |\Gamma (x) \cap \Gamma (y) |,
\end{equation}
where  $|.|$ is the cardinality of the set and $\Gamma (x)$ is the set of neighbours of node $ x $.

\item[] \textbf{Adamic Adar (AA)~\cite{Adamic:2003}:} Modification of CN link prediction method which assigns more weights to the less-connected common neighbors. In this case, the score is given by
\begin{equation}
AA(x,y)= \sum_{z\in (\Gamma(x)\cap \Gamma (y))}\frac{1}{\log|\Gamma (z)|},
\end{equation}
where $z$ runs over the common neighbours of $x$ and $y$.

\item[] \textbf{Resource Allocation (RA)}~\cite{Zhou:2009RA}: This local link prediction method is based on resource allocation dynamics in complex networks~\cite{Qu:2007}. It measures the similarity between unconnected nodes $x$ and $y$ by the number of resources that node $y$ can receive from node $x$. To that end, the common neighbours of the given nodes are considered as transmitters, which distribute a unit of resource equally among all of their neighbours. Therefore, the RA similarity index is given by
\begin{equation}
RA(x,y)= \sum_{z\in (\Gamma(x)\cap \Gamma (y))}\frac{1}{|\Gamma (z)|}.
\end{equation}

\item[] \textbf{Cannistraci-Hebb (CH) network automata model~\cite{ Cannistraci:2017CH}:} Local and parameter-free link prediction method which considers local-communities to compute the similarity scores in a given network. The local-community for each pair of nodes is defined as the subgraph consisting of the common neighbours of the nodes under analysis and the links among them. For each pair of unconnected nodes $x$ and $y$, the CH score is computed as
\begin{equation}
CH(x,y) = \sum_{z \in (\Gamma (x) \cap \Gamma (x))} \dfrac{|\varphi(z)|}{|\Gamma (z)|},
\end{equation}
where $ \varphi(z) $ is a subset of neighbours of $z$ which also belong to the common neighbours of $x$ and $y$. Therefore, this method not only employs the set of common neighbours, but also considers the links between them to compute the similarity indices.

\item[] \textbf{Structural Perturbation Method (SPM)~\cite{Lu:2015}:} This global link prediction method assumes that the structural features of a given network will be consistent before and after perturbing its adjacency matrix. A fraction $p^{H}$ of links are randomly selected to partition the links of the network into a perturbation set, $\Delta E$, and remaining set, $E^R$. The adjacency matrix of the network is, therefore, $A= A^R + \Delta A$, where $A^R$ and $\Delta A$ are the adjacency matrices of the remaining and the perturbation sets, respectively. The links of $\Delta E$ are used to perturb $A^R$. Applying the first-order approximation, the perturbed matrix can be calculated by keeping fixed the eigenvectors of $A^R$, $x_k$, and correcting its eigenvalues, $\lambda_k$, as $\lambda_k+ \Delta\lambda_k$, where $\Delta\lambda_k$ are computed as
\begin{equation}\label{eq:spm1}
\Delta\lambda_{k} \approx \frac{x_{k}^{T} \Delta A x_k }{{x_{k}}^T x_k}.
\end{equation}
Finally, the perturbed matrix is given by
\begin{equation}\label{eq:spm2}
\tilde{A}=\sum_{k=1}^{N} (\lambda_k + \Delta \lambda_k) x_k x_{k}^{T},
\end{equation}
where $N$ is the number of nodes in the network. By taking the average of the perturbed matrices over independent perturbation sets $\Delta E$, we obtain the matrix $\langle \tilde{A} \rangle$, the elements of which corresponding to non-observed links yield the required scores to be used for link prediction. In this work, we set $p ^{H}$ to 0.1.

\item[] \textbf{Fast probability Block Model (FBM)~\cite{liu2013correlations}:} This global link prediction method is based on the general idea of partitioning the nodes of a complex network into different groups and computing the similarity of unconnected nodes by considering the groups to which they belong. To sample a subset of relevant partitions from all possible ones, FBM employs a greedy method in which the network is randomly divided into two blocks. For each block, the subset of nodes that form the largest clique are iteratively removed and create a new block. Finally, the subset of remaining nodes that do not belong to any clique form another block together. After sampling 50 partitions, the FBM score is computed as
\begin{equation}
FBM(x,y) = \frac{1}{|P|} \sum_{p \in P} F(g_x , g_y),
\end{equation}
where
\begin{equation}
    F(\alpha, \beta)= \left\lbrace
\begin{aligned}
    \frac{r_\alpha}{2r_\alpha - l_\alpha}, \, & \alpha=\beta \\    
   \frac{l_{\alpha \beta}}{r_{\alpha \beta} + l_{\alpha \beta} },\,     &  \alpha \neq\beta
\end{aligned}\right.
\end{equation}
In the above expressions, $P$ is the set of sampled partitions. For each partition $p$, $g_x$ and $g_y$ are the particular blocks to which $x$ and $y$ belong. $l_\alpha$ is the number of links between the nodes belonging to block $\alpha$ and $r_\alpha$ is the maximum number of possible links among them. $l_{\alpha \beta}$ is the number of links between blocks $\alpha$ and $\beta$ and $r_{\alpha \beta}$ is maximum number of possible links among them.    

\end{itemize}

\section{Theoretical OS curve}\label{app:tos}
Let ensemble $\mathcal{E}_N$ be characterised by the set of connection probabilities $\lbrace p_l \rbrace$, where index $l$ runs over all possible pairs of $N$ nodes such that $p_l \geq p_{l+1},\, \forall l$. If links are removed with probability $q$, the probability for an edge $l$ not to belong to $G_{\mathrm{obs}}$ is given by the sum of the probabilities for it not to belong to $G$, $P\left( a_l = 0, a_l^{\mathrm{obs}} = 0 \right) = 1 - p_l$, and for it to belong to $G$ and being randomly removed, $P\left( a_l = 1, a_l^{\mathrm{obs}} = 0 \right) = qp_l$, that is,
\begin{equation}
P\left( a_l^{\mathrm{obs}} = 0 \right) = 1 - p_l + qp_l.
\end{equation}
Now, in order to compute the expected precision of the Optimal Strategy, the basic idea is to compute the expected number of correct predictions, $T$, when following the ranking of probabilities until the expected number of non-observed links, $H$, matches the expected number of missing links, $L = q \sum_l p_l$. Hence, we initialise both $T$ and $H$ to zero and, for every link $l = 1, \ldots$, we update them as
\begin{equation}
T_{\mathrm{new}} = T_{\mathrm{old}} + P\left( a_l = 1, a_l^{\mathrm{obs}} = 0 \right) = T_{\mathrm{old}} + q p_l
\end{equation}
and
\begin{equation}
H_{\mathrm{new}} = H_{\mathrm{old}} + P\left( a_l^{\mathrm{obs}} = 0 \right) = H_{\mathrm{old}} + 1 - p_l + qp_l.
\end{equation}
When $H \approx L$, the precision can be computed as $\langle Q \rangle = T/H$. Notice that, for the ER model, $p_l = p, \, \forall l$, and hence the expected precision is $\langle Q \rangle = \frac{qp}{1-p+qp}$, which agrees with the exact value.

\section{Structural consistency index}\label{app:sci}
The Structural Consistency index~\cite{Lu:2015} measures the intrinsic predictability of a given network. The calculation of this index is similar to SPM but, instead of being applied on an incomplete network to predict its missing links, it is applied on the complete network. To that end, a fraction of links is randomly selected from the complete network with adjacency matrix $A$ and set of links $E$ to construct the perturbation set ($\Delta E$ and $\Delta A$). The remaining links define $E^R$ and $A^R $ through the relations $E = E^R + \Delta E$ and $A = A^R + \Delta A$. The perturbed matrix $\tilde{A}$ is then computed using Eqs.~\eqref{eq:spm1} and \eqref{eq:spm2}, respectively. All the links in $U-E^R$ are ranked in descending order based on their scores in $\tilde{A}$, where $U$ is the universal set of links. Finally, the top-L links $E^L$ are selected to compute the structural adjacency index as
\begin{equation}
\sigma_c= \frac{|E^L \cap \Delta E|}{|\Delta E|},
\end{equation} 
where $L=|\Delta E|$.

\section{Details of network ensemble models} \label{app:c}

\begin{itemize}

\item[] \textbf{The soft Configuration Model (sCM):} In the soft Configuration Model~\cite{Park:2003}, each node $i$ is assigned an expected degree $\kappa_{i}$, and each pair of nodes $i$ and $j$ is connected according to the ensemble connection probabilities given by
\begin{equation}\label{eq:conn_prob_CM}
p_{ij} = \frac{\mu \kappa_{i}\kappa_{j}}{1+\mu \kappa_{i}\kappa_{j}},
\end{equation}
with $\mu$ a free parameter controlling the number of resulting edges in the network. If one takes $\mu=1/(\langle k \rangle N)$, then the degree of every node $i$ in the generated networks, $k_i$, is approximately its expected degree, $k_i \approx \kappa_{i}$. 

Given the degrees $k_{i}^{\mathrm{obs}}$ in an observed graph which has been produced by removing a fraction $q$ of nodes from a complete graph in the CM ensemble, we can estimate the expected degrees and the connection probabilities in the complete graph from Eq.~\eqref{eq:conn_prob_CM} as
\begin{equation}\label{eq:conn_prob_CM_proxy}
\tilde{p}_{ij} = \frac{\frac{k_{i}^{\mathrm{obs}}k_{j}^{\mathrm{obs}}}{(1-q)\langle k^{\mathrm{obs}} \rangle N}}{1+\frac{k_{i}^{\mathrm{obs}}k_{j}^{\mathrm{obs}}}{(1-q)\langle k^{\mathrm{obs}} \rangle N}}.
\end{equation}

\item[] \textbf{The geometric $\mathbb{S}^1$ network model:} In the $\mathbb{S}^1$ model~\cite{Serrano:2008ga}, every node $i$ is characterized by a hidden degree and an angular coordinate $(\kappa_i,\theta_i)$ representing the popularity and similarity dimensions. The angular coordinate is distributed at random in similarity space, which is taken to be a one-dimensional sphere, or circle, of radius $R$ adjusted to have a density of nodes equal to 1. Every pair of nodes is connected with a probability 
\begin{align} \label{eq:probability_of_connection_S1}
  p_{ij} = \frac{1}{1 + \left( \frac{R \Delta\theta_{ij}}{\mu \kappa_i \kappa_j} \right)^\beta},
\end{align}
where $\Delta\theta_{ij}$ stands for the angular separation between the nodes in the similarity circle, and the parameters $\mu$ and $\beta$ control the average degree of the network and its level of clustering, respectively. In the limit of $N \rightarrow \infty$, and for large degrees, the expected degree $\langle k_i \rangle$ of a node $i$ in the generated network is its hidden degree $\langle k_i \rangle = \kappa_{i}$.

\item[] \textbf{The degree-corrected Stochastic Block Model (dc-SBM):} In the dc-SBM model~\cite{karrer2011}, each node $i$ is assigned an expected degree $k_i$ and a group $g_i$ determining the community to which it belongs, which is chosen in an arbitrary way. Then, parameter $\theta$ for every node $i$ is computed as
\begin{align} \label{eq:Tetha_dc-SBM}
 \theta_i = \frac{k_i}{\kappa_{g_i}},
\end{align}
where $\kappa_{g_i}$ is the sum of the degrees of all the nodes in group $g_i$. Therefore, each group $g$ fullfils the constraint  
\begin{align} \label{eq:Tetha_Norm}
\sum_{i \in g} \theta_{i}  =1.
\end{align}
Finally, $\omega$ is a matrix of size $K \times K$ controlling the number of links between pairs of groups, where $K$ is the total number of groups. Each element of the matrix is calculated as
\begin{align} \label{eq:Omega_dc-SBM}
\omega_{rs}= \lambda \omega_{rs} ^ {planted} + (1- \lambda) \omega_{rs} ^ {random},
\end{align}
where $\omega_{rs} ^ {random}$ corresponds to a random network with specific expected degree sequence, $\omega_{rs} ^ {random}= \kappa_r \kappa_s / 2m $, where $m$ is the total number of links in the network. On the other hand, $\omega_{rs} ^ {planted}$ generates group structure. For example, in a network with four groups, this matrix is given by

\begin{align}
\omega^ {planted}=
   \begin{bmatrix} 
    \kappa_1 & 0 & 0 & 0 \\
   0 & \kappa_2 & 0 & 0\\
   0 & 0 & \kappa_3 & 0 \\
   0 & 0 & 0 & \kappa_4 \\
   \end{bmatrix} .
\end{align}
 
When $\lambda=0$ links are placed among pairs of nodes at random considering the degree sequence, while when $\lambda=1$links are located within communities. Therefore, any other values for $\lambda$ will result in a combination of the above extremes.

In the dc-SBM model, the number of links placed among two nodes $i$ and $j$ follows a Poisson distribution with mean $ \theta_i \theta_j \omega_{g_i, g_j}$.
However, in the sparse-network limit, the probability for multi-edges to occur is generally low, so $ \theta_i \theta_j \omega_{g_i, g_j}$ is simply taken to be the connection probability. Since these amounts can be larger than 1, in this work, we consider
\begin{equation} \label{eq:Conn_Pro_dc-SBM}
p_{ij} = \frac{\theta_i \theta_j \omega_{g_i, g_j}}{1 + \theta_i \theta_j \omega_{g_i, g_j}}.
\end{equation}

\end{itemize}

\section{Description of real networks datasets.} \label{app:d}

\begin{itemize}
\item[] \textbf{Music} ~\cite{Music:2012}: The nodes in the Music network represent codewords extracted for every single chord in a large set of songs, and directed links connecting consecutive codewords represent transitions among them. To sparsify the network, the disparity filter~\cite{serrano2009extracting} is applied  with parameter $\alpha=0.01$. Finally, we consider an undirected version of network by replacing bidirectional links with undirected ones.  

\item[] \textbf{Drosophila}~\cite{Drosophila:2013}: Nodes represent neurons within the Drosophila optic medulla and links represent fiber tracts connecting neurons.

\item[] \textbf{WTW} ~\cite{WTW:2016}: Backbone of the international trade network in 2013, where nodes represent countries and links are placed among significant trade partners. 

\item[] \textbf{Internet} ~\cite{Internet:2009}: Internet topology at the level of Autonomous Systems (AS) level corresponding to June 2009 and collected by the Cooperative Association for Internet Data Analysis (CAIDA). We removed nodes with degree lower than 5 to produce a reduced size version.

\item[] {\textbf{Florida Food Web}}~\cite{FFW:2005}: Food web in the Florida Bay ecosystem, in which every directed link connects a prey to its predator. We consider the undirected version of this network created by placing an undirected link between every pair of nodes connected by at least a single directed link.

\item[] \textbf{Word Adjacency}~\cite{WordAdjacency:2006}: Adjacency network where nodes represent a selected set of common nouns and adjectives in the novel of David Copperfield by Charles Dickens, and links are placed between adjacent pairs of words in the book.

\item[] \textbf{Karate}~\cite{Karate:1977}: Social network of a karate club members where each link connects a pair of members who communicate outside the club.
 
\item[] \textbf{Polbooks}~\cite{Polbooks:2006}: Nodes of this network represent the books on the topic of the US politics and links represent the pairs of books bought on Amazon by the same customers.  

\end{itemize}

\end{document}